\documentclass[superscriptaddress,aps,pra,twocolumn,preprintnumbers,amsmath,amssymb,floatfix,groupedaddress]{revtex4}

\usepackage{enumerate}
\usepackage{graphicx}
\usepackage{color}

\newcommand{\beq}{\begin{equation}}
\newcommand{\eeq}{\end{equation}}
\newcommand{\bea}{\begin{eqnarray}}
\newcommand{\eea}{\end{eqnarray}}
\newcommand{\ba}{\begin{array}}
\newcommand{\ea}{\end{array}}

\newcommand{\bef}{\begin{figure}}
\newcommand{\eef}{\end{figure}}

\begin{document}

\title{A simple derivation of Born's rule with and without Gleason's theorem.}
\author{Alexia Auff\`eves$^{(1)}$ and Philippe Grangier$^{(2)}$} 
\affiliation{
(1): Institut N\' eel$,\;$BP 166$,\;$25 rue des Martyrs$,\;$F38042 Grenoble Cedex 9, France. \\
(2): Laboratoire Charles Fabry, Institut d'Optique, CNRS, Univ.~Paris~Sud, \\
2 avenue Augustin Fresnel, F91127 Palaiseau, France.
 }

\begin{abstract}

We present a derivation of Born's rule and unitary transforms in Quantum Mechanics, from a simple set of axioms built  upon a physical phenomenology of quantization. Combined to Gleason's theorem, this approach naturally leads to the usual quantum formalism, within a new conceptual framework that is discussed  heuristically in details. The structure of Quantum Mechanics, from its probabilistic nature to its mathematical expression, appears as a result of the interplay between the quantized number of  ``modalities"  accessible to a quantum system, and the continuum of  ``contexts" that are required to define these modalities. 

\end{abstract}


\maketitle

\noindent {\bf 1. Introduction.} 

\vskip 3mm

Deriving Born's rule, rather than postulating it as it is done in standard textbooks, has been  
envisioned since the early times of Quantum Mechanics (QM) \cite{Born}. 
A major asset in this direction is  Gleason's theorem  \cite{gleason}, whose critical importance for the foundations of QM has been recognized since it was published in 1957. The theorem is simple to state (see below), but difficult to demonstrate, and a nice presentation is provided in \cite{helena}.  

The main attempts to use Gleason's theorem for deriving Born's rule, and then the whole quantum formalism, have been done in the framework of formal quantum logic \cite{Piron}.  However, such approaches were not  considered very appealing by physicists, and though Gleason's theorem essentially gives the correct quantum probability law, it was often said that it provides no physical insight into why the result should be regarded as probabilities. According to \cite{zurek}, it is even considered as a motivation to seek a more physically transparent derivation of Born's rule.
This is partly because the hypothesis of Gleason's theorem do not fit easily in the usual ``wave function" approach of QM, and in particular within the superposition principle, which is usually put forward as the very first postulate when introducing quantum mechanics. 

In this  paper, we will introduce  new axioms for QM \cite{r1,r2,r3z,r3,r4,r5,r6,r7,r8}, starting with three physical axioms defining quantum rules, without any mathematical formalism. When completed by a fourth mathematical axiom, it will turn out that the four together correspond to the hypothesis of Gleason's theorem, leading straightforwardly  to Born's rule.  Before stating this fourth axiom, we 
will introduce it heuristically, by using the three physical ones, completed by a set of physical assumptions. 
So let us start with  the following set of physically motivated axioms, which have been introduced and discussed  in  \cite{ph1,ph2,csm1}:
\begin{itemize}

\item {\bf Axiom 1} (quantum ontology): 
Given a physical system, a {\it modality} is defined as the values of a complete set of physical quantities  that can be predicted with certainty and measured
repeatedly on this system. 
The complete set of physical quantities  is called a ``context", and the 
modality  is attributed {\bf jointly}  to the system and the context. 

\item {\bf Axiom 2} (quantization):  For a given context, that is a given ``knob settings" of the measurement apparatus,
there exist $N$ distinguishable modalities $\{u_i\}$, that are mutually exclusive~: if one is true, or verified, 
the other ones are wrong, or not verified. The value of $N$, called the dimension, is a characteristic 
property of a given quantum system.

\item {\bf Axiom 3} (changing contexts):  The different contexts relative to a given quantum system are related between themselves 
by (classical) transformations $g$ that have the structure of a continuous group ${\cal G}$. 

\end{itemize}

For the sake of clarity, we  note
that, within the usual QM formalism (not used so far), a modality and a context correspond
respectively to a pure quantum state, and to a complete set of commuting observables. 
The axioms are formulated for a finite $N$, but this restriction will be lifted below.  
These axioms, under the acronym ``CSM", meaning Context, System, Modality,
have been discussed in \cite{csm1}, both physically and philosophically, 
and we will not reproduce this discussion here.  
We will rather consider  the following question:
it is  postulated in Axiom 2  that there are $N$ mutually exclusive modalities 
associated to each given context,   but there are many more modalities, 
corresponding to all possible contexts. These modalities are generally not mutually exclusive, 
but are {\bf incompatible}: it means that if one is true, one cannot tell whether the other one  is true or wrong.
Then, how to relate between themselves all these modalities ? 
 
A first crucial result already established  in \cite{csm1} is that this connection can only be a probabilistic one,  otherwise the axioms would be  violated; the argument is as follows.
Let us consider a single system, two different contexts $C_u$  and $C_v$, and the associated modalities $u_i$ and $v_j$, where $i$ and $j$ go from 1 to $N$. The quantization principle (Axiom 2) forbids to gather all the modalities $u_i$ and $v_j$ in a single set of $2N$ mutually exclusive modalities, since their number is bounded to $N$. Therefore the only relevant question to be answered by the theory is: If the initial modality is $u_i$ in context $C_u$, what is the {\it conditional probability} for obtaining modality $v_j$ when the context is changed from $C_u$ to $C_v$ ? We emphasize that this probabilistic description is the unavoidable consequence of the impossibility to define a unique context making all modalities mutually exclusive, as it would be done in classical physics. It appears therefore as a joint consequence of the above Axioms 1 and 2, i.e. that modalities are quantized, and require a context to be defined. 

Now, according to Axiom 3, changing the context results from changing the measurement apparatus 
at the macroscopic level, that is, ``turning knobs". A typical example is changing
the orientations of a Stern-Gerlach magnet. These context transformations have the mathematical structure of a continuous
group, denoted ${\cal G}$~: the combination of several transformations is associative  and gives a new transformation,
there is a neutral element (the identity), and each transformation has an inverse. 
Generally this group is not commutative : for instance, the three-dimentional
rotations associated with the orientations of a Stern-Gerlach magnet do not commute.
For a given context, there is a given set of $N$ mutually exclusive modalities, denoted $\{u_i\}$. 
By changing the context,  one obtains $N$ other mutually exclusive modalities, denoted $\{v_j\}$, 
and one needs to build up a mathematical formalism, able  to provide  
the probability that a given initial modality $\{u_i\}$ ends up in a new modality $\{v_j\}$.

The standard approach at this point is to postulate that each modality $u_i$  is associated with a
vector $|u_i \rangle$  in a $N$-dimensional Hilbert space, and that the set of $N$  mutually 
exclusive modalities in a given context  is associated to a set of $N$ orthonormal vectors.
Rather than  vectors $|u_i \rangle$ and $|v_j \rangle$,  one can 
equivalently use rank-one projectors $ P_{u_i} $ and $P_{v_j}$, and 
Born's rule giving the conditional probability $p(v_j | u_i)$ can be written as  
\beq
p(v_j | u_i) = \mathbf{Trace}( P_{u_i} P_{v_j}).
\label{born1}
\eeq

As we will  show below, after postulating
that each modality  is associated to an Hermitian projector acting in a suitable Hilbert space, 
 there is actually no need to postulate also Born's rule: it follows immediately as a consequence of Gleason's theorem, 
and the transformation of  projectors  associated with a change of context must be unitary. 
Before doing that, we shall  first  justify heuristically why each modality  should be associated to a projector.
This will be done in Section 2, then we will come back to Gleason's theorem in Section 3,
and finally discuss some consequences of our approach in Section 4. 
\\
\\
{\bf 2. Heuristics without Gleason's theorem.}
\\
\\
The main goal of this heuristics is 
to give a justification for Axiom~4 (given in Section 3 below), telling 
that each modality  is associated to an Hermitian projector acting in a suitable Hilbert space. 
For this purpose, we will start from Axioms 1-3 only, and introduce a set of assumptions
to construct a consistent probability theory,  
by imposing some requirements on what it should describe. This will lead us 
to associate modalities with projectors  in an Hilbert space, and to
get Born's rule and unitary transforms on the way.  The more formal proofs will 
be given in Section 3 by using Axiom 4.

\noindent{\it The general probability matrix}
\vskip 2mm

Since the $\{u_i\}$ and $\{v_j\}$ are by definition non-exclusive modalities,
one has to introduce the probabilities of finding the particular modality
$v_j$ (in the new context), when one starts in modality $u_i$  (in the old context).
There are $N^2$ such probabilities, 
that  can be arranged  in a matrix $\Pi_{v|u} = \left( p_{v_j|u_i} \right)$, containing
 all  probabilities  connecting the $N$ modalities in each context  $\{u_i\}$ and $\{v_j\}$. 
Since one has obviously $0 \leq p_{v_j|u_i}  \leq 1$ and 
$\Sigma_j \; p_{v_j|u_i} = 1$, the matrix $\Pi_{v|u}$ is said to be a {\it stochastic} matrix. 

For clarity, let us emphasize  the interpretation of the conditional probability notation:
in agreement with the definition of modalities as certainties, 
 the meaning of $p_{v_j|u_i}$ is  that ``if we start  (with certainty) from modality $u_i$
 in the old context, then the probability to get modality $v_j$ in the new context is $p_{v_j|u_i}$". 
 The matrix of all $p_{v_j|u_i}$ provides the starting point for our heuristic approach, 
by which theoretical predictions are connected to experiments.
For $N=3$, one will have for instance
$$\Pi_{v|u} = \left(  \begin{array}{lll} 
 p_{v_1|u_1}, & p_{v_2|u_1}, & p_{v_3|u_1}  = 1 -  p_{v_1|u_1}- p_{v_2|u_1}\\ 
 p_{v_1|u_2}, & p_{v_2|u_2}, & p_{v_3|u_2}  = 1 -  p_{v_1|u_2}- p_{v_2|u_2} \\ 
 p_{v_1|u_3}, & p_{v_2|u_3}, & p_{v_3|u_3}  = 1 -  p_{v_1|u_3}- p_{v_2|u_3}  \end{array} \right)$$
As we will see below, $N \geq 3$ is required because some crucial properties 
of  $\Pi_{v|u} $ do not show up for $N=2$.

Let us also define a ``return"
probability matrix $\Pi_{u|v}$, by exchanging the roles of the initial and final contexts. 
The matrix $\Pi_{u|v}$ has the same properties as $\Pi_{v|u}$, but these two matrices 
are {\it a priori} unrelated, whereas it is well known that in standard QM, they are transpose 
of each other. In the following, we will  introduce 
simple assumptions which will constraint  these matrices
to being  {\it unistochastic}, i.e. that their coefficients are the square 
moduli of the coefficients of a unitary matrix \cite{unis}; and then, to being  transpose 
of each other.
\\

\noindent{\it A mathematical identity}
\vskip 2 mm

In order to manipulate the $\Pi_{v|u}$ and $\Pi_{u|v}$ matrices,  it is convenient to  introduce
orthogonal ($N \times N$) projectors $P_i$, that are zero everywhere, except for the $i^{th}$ 
term on the diagonal that  is equal to 1. These projectors verify the relation $P_i P_j = P_i \delta_{ij}$.
A useful operation is then to extract the particular probability $p_{v_j|u_i}  $ from  $\Pi_{v|u}$, 
or $p_{u_i | v_j}  $ from  $\Pi_{u|v}$, and
one has the following identities~:
\bea
p_{v_j|u_i} &=&\mathbf{Tr}( P_j \;  \Sigma_{v|u}^\dagger \;   P_i \;  \Sigma_{v|u}) = \mathbf{Tr}(P_i \;  \Sigma_{v|u} \; P_j \;  \Sigma_{v|u}^\dagger)\; \; \; \; \; \;   \label{trace1} \\Ê
p_{u_i|v_j} &=&\mathbf{Tr}(P_i \;  \Sigma_{u|v}^\dagger \; P_j \;  \Sigma_{u|v}) = \mathbf{Tr}(P_j \;  \Sigma_{u|v} \; P_i \;  \Sigma_{u|v}^\dagger)\; \; \;  \; \; \;   \label{trace2}
\eea
where $\mathbf{Tr}$ is the Trace, $^\dagger$ is the Hermitian conjugate, and 
\beq \Sigma_{v|u} = \left[ e^{i \phi_{v_j|u_i} } \sqrt{p_{v_j|u_i}  } \right],  \; 
\Sigma_{u|v} = \left[ e^{i \phi_{u_i|v_j} } \sqrt{p_{u_i|v_j}  } \right] \eeq
are $N \times N$ matrices  formed by square roots of the probabilities, and by arbitrary phase factors which are introduced for the sake of generality, 
and don't play any role at that stage.  
We  emphasize that the equations above 
are only mathematical identities, and don't tell more than
what is already contained in the definition of the matrices $\Pi_{v|u}$ and $\Pi_{u|v}$.
A useful  marginal case is the situation where the context is not changed, so $u \equiv v$, and 
\beq
p_{u_j|u_i}   = \mathbf{Tr}(P_j \; P_i)= \delta_{ij}.
\label{mut}
\eeq
where $p_{u_j|u_i}  =  \delta_{ij}$ is obviously consistent with mutually exclusive modalities within  a given context.

From Eqs. \ref{trace1}, \ref{trace2} the elements $p_{ji}$ of a general stochastic matrix $\Pi$ can  be written as
(the subcripts ${u|v}$ or ${v|u}$ are omitted for simplicity):
\beq p_{ji} = \mathbf{Tr}(P_i \;  \Sigma \; P_j \;  \Sigma^\dagger). \label{pji} \eeq
Now, according to the singular values theorem, there must exist two unitary matrices $U$ and $V$, 
and a real diagonal matrix $R$, such that 
\beq \Sigma  = U \, R \; V^\dagger, \; \; \; \; \Sigma^\dagger = V \, R \; U^\dagger \label{svdm} \eeq
where the diagonal values of $R$ are the square roots of the (real)  eigenvalues of  
$\Sigma \Sigma^\dagger$, equal to those of  $\Sigma^\dagger  \Sigma$, and are called the singular values of $\Sigma$ \cite{dem-svd}. The matrix $\Sigma$ is unitary iff $R$ is the identity matrix $\hat 1$.
We note that the value of $ \mathbf{Tr}(R^2)$
is  the sum of the square moduli of all the coefficients of $\Sigma$, and is therefore equal to $N$.
 For a generic stochastic  matrix $\Pi$, $\Sigma \Sigma^\dagger$ has diagonal coefficients equal to 1, 
but is not diagonal, whereas $R^2$ is diagonal,  and its $N$ coefficients are real, positive, and sum to $N$, but are not necessarily equal to one.

Using Eqs. (\ref{pji}, \ref{svdm}), $p_{ji}$ can now be written:
\bea
p_{ji} &=& \mathbf{Tr}(P_i  \;  U R V^\dagger \; P_j \;  V R U^\dagger) \nonumber \\
&= & \mathbf{Tr}\left( (U^\dagger P_i  \;  U) \;   R \; (V^\dagger  P_j   V)  \;  R \right)
\label{svd}
\eea
This equation is again a mathematical identity, on which we shall now impose physical constraints. 
In the section below we will consider  $\Sigma_{v|u}$, but obviously the same arguments are also 
valid  for  $\Sigma_{u|v}$. 
\\

\noindent{\it Physical constraints on the probability matrix.}
\vskip 1mm
Given Axioms 1 and 2, our main physical 
argument is that the probability
$p_{v_j|u_i}$ should only depend on the particular modalities $u_i$ and $v_j$ being considered, and not on the whole contexts in which they are embedded.
This  important property of ``non-contextuality" for modalities   \cite{nonc} 
is related to Gleason's theorem, and it will appear again in Section  3.
It  tells that the same modality can pertain to different contexts, and therefore can be defined (in particular, mathematically) independently of  other modalities in a given context. 
This (quantum) non-contextuality is fully compatible with contextual objectivity  \cite{ph1,ph2,csm1}Ê: the latter states that a modality needs a context to be defined, whereas the former  tells that the same modality can be defined in several contexts. 

In order to fulfill this condition, the decomposition of Eq. (\ref{svd}) suggests that it might be possible 
to separate two parts (within parenthesis) associated with the two specific modalities $u_i$ and $v_j$. 
However, if  the singular values of the matrix  $\Sigma_{v|u}$  are not equal to 1, the matrix $R \neq \hat 1$
will impose a context-dependent structure on the  whole sets of modalities $\{ u_i \}$ and  $\{ v_j \}$. 
There is nevertheless a way to warrant that  $R$ does
not depend on  $\Sigma_{v|u}$, still satisfying the constraints spelled out above: it is to impose that $R = \hat 1$. 
Therefore, in order to have the probability depending on separate mathematical objects associated with each modality,
we will  posit the basic assumption:
\begin{itemize}
\item  Assumption 1: 
In order to ensure that $p_{v_j | u_i} $  depends only on the two particular modalities being considered, 
the $N$ singular values of $\Sigma_{v|u}$ must be all equal together, and thus are all equal to one.
\end{itemize} 

Then as said above $\Sigma_{v|u}$ will be a unitary matrix $U V^\dagger$, but 
one may wonder whether  orthogonal (real) matrices might be enough. 
In order to justify that the full unitary set  is required, we shall use a second assumption:
\begin{itemize}
\item Assumption 2: Since the change of contexts corresponds to a continuous group (Axiom 3), the set of 
matrices $\Sigma_{v|u}$ must be connected in a topological sense, and must contain the identity matrix. 
\end{itemize}

Then it is known that the set of  orthogonal matrices is topologically disconnected 
 in two parts with determinant $+1$ and $-1$, which contradicts the above assumption. For instance, permutation matrices 
are not connected to the identity, whereas they correspond simply to ``relabelling" the modalities, 
i.e. to a trivial change of context. On the other hand, all (complex) unitary matrices are connected to the identity, 
and do agree with Assumption 2 (for other arguments  see refs.  \cite{complex,david,aaronson}). 
\\

\noindent{\it Unitary matrices and Born's formula}
\vskip 2mm

We are thus lead to the conclusion that
$\Sigma_{v|u} $ must be a unitary matrix $S_{v|u} $, 
with $S_{v|u} ^\dagger = S_{v|u}^{-1}$.
Then Eqs. (\ref{trace1}) for picking up a particular probability become:
\begin{eqnarray}
p_{v_j|u_i} &=& \mathbf{Tr}( P_j \,.\,   S_{v|u}^\dagger \,.\,  P_i \,.\, S_{v|u})  \nonumber \\
&=& \mathbf{Tr}(  P_i \,.\, S_{v|u} \,.\,P_j \,.\,  S_{v|u}^\dagger ) 
\label{equ2}
\end{eqnarray}
which shows that  the matrix $\Pi_{v|u} $ must be {\it unistochastic}, 
i.e. made by the square modulus of the coefficients of a unitary matrix. 
Such matrices are also  {\it bistochastic}, i.e. their  lines and rows sum to 1 \cite{bist3}. 
Then we can define 
 \bea 
 P_i' =  \; S_{v|u}^\dagger \, . \,  P_i  \, . \,   S_{v|u}\; \; , \; \; \; P_j''=  \; S_{v|u} \, . \,  P_j  \, . \,   S_{v|u}^\dagger  \; \;   \label{projP}.
 \eea
 It is clear that these operators are all Hermitian projectors, i.e.  one has $P^\dagger = P$ and $P^2 = P$ for each of them,
 and also that all sets $\{ P_i' \}$ and $\{ P_j'' \}$
have the same orthogonality properties as  the initial set of projectors $\{ P_i \}$,
given by Eq. (\ref{mut}). 
One can thus rewrite Eq. (\ref{trace1}) as:
\bea
p_{v_j|u_i} =  \mathbf{Tr}(P_j \;  P_i' ) = \mathbf{Tr}(P_i \;  P_j'' ).  \label{equP} 
\eea
This is just  Born's formula (Eq. \ref{born1}), which is 
obtained here heuristically, rather than from a postulate.

Finally, the obvious next step is
to associate projectors with modalities in  each context,
and for the matrix $\Pi_{v|u} $ it can be done in two consistent ways as seen above:
\bea
\text{old context} \{u_i \}&\rightarrow& \text{new context} \{v_j \}   \\
P_i' =  \; S_{v|u}^\dagger \, . \,  P_i  \, . \,   S_{v|u} &\rightarrow& P_j \nonumber \\
P_i  &\rightarrow& P_j''=  \; S_{v|u} \, . \,  P_j  \, . \,   S_{v|u}^\dagger \nonumber 
\label{Pi}
\eea
One can now come back to the matrix $\Pi_{u|v}$, for which the same reasoning is valid, 
and leads to a unitary matrix $S_{u|v}$. By reverting the contexts one has thus:
\bea
\text{old context} \{v_j \}&\rightarrow& \text{new context} \{u_i \}  \\
Q_j''=  \; S_{u|v}^\dagger \, . \,  P_j  \, . \,   S_{u|v} &\rightarrow& P_i \nonumber \\
P_j &\rightarrow& Q_i' =  \; S_{u|v} \, . \,  P_i  \, . \,   S_{u|v}^\dagger  \nonumber 
\label{Qi}
\eea
But since projectors are now associated with modalities, 
they should be the same for a given modality  in a given context, 
i.e. one should have $P_j''=Q_j''$, and $P_i' =Q_i' $. This is obtained if $S_{u|v}$ is the inverse of $S_{v|u}$,
leading to a third assumption:
\begin{itemize}
\item  Assumption 3: 
In order to associate projectors with modalities in a consistent way, 
the matrices $\Pi_{u|v} $ and $\Pi_{v|u} $ must be related by 
$S_{u|v} = S_{v|u}^\dagger= S_{v|u}^{-1} $, and  thus $\Pi_{u|v} = \Pi_{v|u}^{t}$.
\end{itemize}

Then the various point of views represented in the relations (\ref{Pi}, \ref{Qi})
are all consistent and give the same values for the probabilities, because 
each $S_{v|u}$  can be associated to an element of the group of context transformations ${\cal G}$, and 
its inverse is $S_{u|v} = S_{v|u}^{-1} = \, S_{v|u}^\dagger$. 
For the general consistency of the approach, 
this set of matrices gives a $N \times N$ (projective) representation of the group of context transformations;
this is fully consistent with the well known Wigner theorem \cite{fl}.
This continuous unitary evolution will be essential 
to describe the evolution  of the system (translation in time), and it is also related to Theorem 2 in  Sec. 3 below. 

The  identification of the matrix 
$\Sigma_{v|u} $ as being a unitary matrix $S_{v|u} $, and the association of projectors to modalities, 
are the results we were looking for;  in 
the next section they will be restated as Axiom 4 in our framework. 
In the above heuristic calculation, valid in the finite-dimentional case, they 
appear to be a joint consequence of the three assumptions made above, 
and of the mathematical identity given by Eqs. (\ref{trace1}, \ref{trace2}). 
By starting from Axiom 4, the Trace formula used in this identity
will  turn out to be the only possible choice. 

We also obtained Born's formula,  apparently avoiding the heavy machinery of Gleason's theorem, because we use the tools of linear algebra 
applied to real or complex $N \times N$ matrices, where all the required mathematical properties are already embedded. 
In order to obtain a full mathematical proof, we will now formally state Axiom 4, and deduce Born's rule in the general case. 
\\

{\bf 3.  Born's rule  from Gleason's theorem.}
\\

Here we add explicitly a fourth axiom, which associates modalities with projectors in a Hilbert space.
Then we will demonstrate two theorems, which are respectively Born's rule, and the unitary evolution of projectors. 
The axiom and the theorems are as follows.

\begin{itemize}

\item {\bf Axiom 4 }: For a system with dimension $N$, 
each modality is bijectively  associated with a $N \times N$
Hermitian rank-one projector $P_i$ ($P_i^\dagger = P_i^2 = P_i$).
Each set of $N$ modalities within a given context is associated to a set  of $N$ such projectors, 
verifying $ P_i P_j = P_i  \; \delta_{ij}$,  and $\sum_i P_i = \hat 1$.  
The same projector (and therefore the same modality) may be part of different contexts. 

\item {\bf Theorem 1 (Born's formula)}:
If the system is known to be in the modality  $u_i$ from the set $\{u_i\}$,
the probability that it is found in modality $v_j$ from the set $\{v_j\}$ 
corresponding to another context  obtained by 
the context transformation $g_{v|u}$ is: 
\beq
p_{v_j|u_i} = \mathbf{Tr}(  P_i \,.\, P'_j  ) 
\label{born}
\eeq
where $P_{i}$ and $P'_{j}$  are respectively associated to the 
modalities $u_i$  and $v_j$.

\item{\bf Theorem 2 (unitary transforms)}:
The different sets of projectors corresponding to different contexts are related by 
unitary transformations $S_{v|u}$,  
i.e. one has $P'_j = S_{v|u} \,.\,P_j \,.\,  S_{v|u}^\dagger$.
\end{itemize}
{\it Proof.} Let us first remind Gleason's  theorem \cite{gleason,helena}~:  
\vskip 2mm 

{Ê\it Let $f(P_i)$ be a function of rank-one projectors $P_i$ in a real or complex Hilbert space with a dimension larger  than 2, to the interval [0,1] of real numbers.  Let assume that  $\sum_i f(P_i) = 1$ for any set $\{ P_i \}$ of  
mutually orthogonal projectors ($ P_i P_j = P_i  \; \delta_{ij}$) verifying $\sum_i P_i = Id$. Then there is a unique positive Hermitian operator $\rho$  with unit trace so that $f(P_i) = \mathbf{Tr}(\rho P_i)$ for all $P_i$. }
\vskip 2mm

According to Axiom 4, each modality is  bijectively associated  to an Hermitian rank-one projector $P_i$, and  any context is associated to a set of $N$ mutually orthogonal projectors $\{ P_i \}$ verifying $\sum_i P_i = Id$.  Let us start from a context $\{ P_i \}$, and go to another context $\{ P_j' \}$, which may actually be the same as $\{ P_i \}$. Since one has necessarily $\sum_j p(P_j' | P_i)  = 1$ for any set  $\{ P_j' \}$, there exist a unique $\rho$ such that $p(P_j' | P_i)  = \mathbf{Tr}(P_j'  \rho)$. 

In addition, one may choose  $P_j' = P_i$, and then 
(since $\rho$ is unique) $p(P_i | P_i)  = \mathbf{Tr}(P_i \rho) = 1$. This is possible only if $\rho = P_i$, and we obtain the expected Born's formula $p(P_j' | P_i)  = \mathbf{Tr}(P_j'  P_i)$. This proves Theorem 1. 

In addition, since $\{ P_i \}$ and $\{ P_j' \}$ are sets of projectors onto two orthonormal basis (Axiom 4), there is a unitary transform $S_{j|i}$ such that  $P_j' = S_{j|i} \,.\,P_j \,.\,  S_{j|i}^\dagger $, up to some relabelling of the basis. This proves Theorem 2. 
\vskip 2mm

Let us note that the main hypothesis for Gleason's  theorem,  i.e. that  probabilities sum to one
for any set of mutually orthogonal projectors summing to identity, 
is a  joint consequence of Axiom 2, i.e. that there are  $N$  mutually exclusive modalities in each given context, 
and Axiom 4, i.e. that  mutually orthogonal projectors are associated to these modalities. 
This remark  allows us to lift the restriction on a finite value of $N$: since Gleason's theorem is valid in any dimension, 
Axioms 2 and 4 can also be considered valid for any $N$ \cite{type1}.
This means  also that the  (classical) additivity of  probabilities can be used within a given context \cite{povm}.

Another more implicit hypothesis is that $f(P_i)$ depends only on $P_i$, and not on other (orthogonal) $P_{j \neq i}$ within the given set in $\sum_j P_j = Id$; this property is usually called ``non-contextuality" \cite{nonc}, and we already introduced it as Assumption 1. It means that, given an initial modality, the conditional probability depends on the particular outcome modality considered, and not on other modalities within a given outcome context.  
Though this hypothesis may be considered very strong \cite{saunders}, it fits perfectly with our ``objective" definition of modalities \cite{ph1,ph2,csm1}: though a modality needs a context to be observed, the same modality may appear in different contexts, always associated with the same projector \cite{spins}.  Therefore the {\bf physical} Axioms 1-3, complemented by the mathematical Axiom 4, do allow us to deduce Born's rule from Gleason's theorem. 
\\

{\bf 4. Discussion}
\vskip 1mm

Since we have now reached the starting point of most QM textbooks \cite{cct},  
it should be clear that 
the standard structure of QM can be obtained from the above axioms \cite{time}.
In particular, one can 
associate the $N$ orthogonal projectors $\{ P_i \}$ to the $N$  orthonormal vectors
which  are eigenstates of these projectors up to a phase factor, i.e., to  rays in the Hilbert space.
Similarly, the expected probability law  for the measurement results $\{a_i \}$ will be obtained 
by writing any physical quantity $A$ as as an operator $\hat A = \sum a_i P_i$, 
this is  the usual spectral theorem. 

We emphasize that we do not need
any additional ``measurement postulate", since measurement is already included
in Axiom 1, i.e. in the very definition of a modality (see detailed discussions in \cite{ph1,ph2,csm1}). 
Quantum superposition are certainly there as usual, but they are not spooky 
``dead-and-alive" concepts: they are rather the manifestation of a modality (i.e., a certainty) in
another context. Entanglement  is also present as linear superpositions of tensor product states, 
corresponding to modalities in a ``joint" context.  In a two-particle Bell-EPR experiment \cite{Bell-EPR,Bell-exp}, 
the entangled modality is defined in a joint context (e.g., a singlet state for two spins), and it is incompatible
with a separable modality corresponding to separate measurements. When a measurement 
is done on one side for one particle only, there is no influence or action at a distance, but
the system (still not measured on the other side) may be quite far from
the new context resulting from the partial measurement.
Since a modality requires both a context and a system, one sees that 
it embeds non-local features, corresponding to quantum non-locality, but fully compatible both 
with relativistic causality and with physical realism \cite{csm1}.
\\

The view about the ``classical vs quantum" dilemma 
that emerges from our approach has been discussed in details in  \cite{csm1}.
It does agree with physical realism, 
given that classical objectivity has been replaced by contextual objectivity \cite{ph1,ph2}, 
as expressed by Axiom 1. 
This Axiom takes from EPR 
their definition of ``elements of physical reality"  \cite{epr} based on 
full predictability and reproducibility, and from Bohr
the idea that 
such a physical reality must include ``the very conditions which define the possible types of predictions 
regarding the future behavior of the system" \cite{bohr}, i.e., the context.  
Therefore physical reality does 
not belong any more to the system alone, but  pertains jointly to the Context, System, and Modality (CSM). 
This approach allows one to distinguish clearly between 
the modality, which is basically a real physical phenomenon, or a physical event in the sense of 
probability theory, and the projector, 
which is a mathematical tool for calculating non-classical probabilities. This point of view also provides novel answers to questions 
about the ``reality of the wave-function". 

To conclude, let us emphasize that we discussed a very idealized version of QM, 
based on pure states and orthogonal measurements. Nevertheless, this idealized
version does provide the basic quantum framework,  and
connects the experimental definition of a physical quantity and the measurement results 
in a consistent way, both physically and philosophically  \cite{csm1}.
Adding more refined tools such as density matrices, imperfect measurements, POVM, open systems, decoherence,
is of great practical interest and use, but this will not ``soften"  
the basic ontology of the theory, as it is presented here. 
The present work, deeply rooted in ontology, is thus  complementary to many recent related proposals
 \cite{r1,r2,r3z,r3,r4,r5,r6,r7,r8}.
\\

{\bf Acknowledgements :}
The authors thank Fran\c cois Dubois  and Anthony Leverrier for many discussions and contributions,
and Nayla Farouki for continuous support.


\begin{references}
\bibitem{dem-svd}  
To show this, diagonalize the Hermitian matrix $\Sigma^\dagger \Sigma$, get the real diagonal matrix $R$ and 
the unitary matrix $V$ so that  $V^\dagger \Sigma^\dagger \Sigma V = R^2$.
Then define another unitary matrix $U$ such that $U R =  \Sigma V$, and 
$R U^\dagger =  V^\dagger \Sigma^\dagger  $,   so that $ U^\dagger \Sigma \Sigma^\dagger U = R^2 $.
Then one gets   $\Sigma = U R V^\dagger$ as expected.
\bibitem{nonc} This definition of non-contextuality is used in articles dealing with Gleason's theorem \cite{helena}, 
and it is sometimes called ``measurement non-contextuality" \cite{r8}. In that sense QM is non-contextual, 
without any contradiction with  the Kochen-Specker theorem.
\bibitem{complex} Short reviews of  arguments leading to the conclusion that neither real numbers nor quaternions are adequate for QM are given e.g. in \cite{david,aaronson}. 
\\
\bibitem{povm} It follows that Gleason's theorem using POVM (see e.g. \cite{helena}) is not useful 
for our purpose, which is to associate orthogonal projectors with events which are certain, repeatable, and mutually exclusive
(see Axioms 1 and 2).
\bibitem{spins} Consider e.g.  a system of two spin 1/2 particles, and define $\vec S = \vec S_2  + \vec S_2$. Using standard notations
for coupled and uncoupled basis,  the $|m_1=1/2, m_2=1/2 \rangle$ modality in the context $\{S_{z1},S_{z2} \}$ is the same 
as the $|S=1, m_S=1 \rangle$ modality in the context $\{\vec S^2, S_{z} \}$, though other modalities in the same  two contexts are different.
\bibitem{time} The postulate  on time evolution is not spelled out, but it enters in the same framework, 
by including translations in time in the group ${\cal G}$. For instance,  if  ${\cal G}$ is the Galileo group, 
standard non-relativistic QM can be recovered, including Schr\"odinger's equation \cite{fl}.
\end{references}
\end{document}